\documentclass[twoside]{dis07}
\usepackage[latin1]{inputenc}
\usepackage[dvips]{graphicx,epsfig,color}
\usepackage{wrapfig,rotating}
\usepackage{amssymb,amsmath,array}

\begin{document}
\title{Hard Diffraction and the Color Glass Condensate}

\author{Cyrille Marquet
%
%
\vspace{.3cm}\\
%
RIKEN BNL Research Center, Brookhaven National Laboratory, Upton, NY 11973, USA
%
}

\maketitle

\begin{abstract}
Following the Good-and-Walker picture, hard diffraction in the high-energy/small$-x$
limit of QCD can be described in terms of eigenstates of the scattering matrix off a
Color Glass Condensate. From the CGC non-linear evolution equations, it is then possible to derive the behavior of diffractive cross-sections at small $x.$ I discuss recent results, in particular the consequences of the inclusion of Pomeron loops in the evolution.
\end{abstract}

\section{Parton saturation and hard diffraction}

When probing small distances inside a hadron with a fixed momentum scale 
$Q^2\!\gg\!\Lambda_{QCD}^2,$ one resolves its constituents quarks and gluons. As one increases the energy of the scattering process, the parton densities seen by the probe grow. At some energy much bigger than the hard scale, the gluon density has grown so large that non-linear effects become important. One enters the saturation regime of QCD, a non-linear yet weakly-coupled regime that describes the hadron as a dense system of weakly interacting partons (mainly gluons). 

The transition to the saturation regime is characterized by the so-called saturation momentum $Q_s(x)\!=\!Q_0\ x^{-\lambda/2}.$ This is an intrinsic scale of the high-energy hadron which increases as $x$ decreases. $Q_0\!\sim\!\Lambda_{QCD},$ but as the energy increases, $Q_s$ becomes a hard scale, and the transition to saturation occurs when $Q_s$ becomes comparable to $Q.$ Although the saturation regime is only reached when $Q_s\!\sim\!Q,$ observables are sensitive to the saturation scale already during the approach to saturation when $\Lambda_{QCD}\!\ll\!Q_s\!\ll\!Q.$ This is especially true in the case of hard diffraction in deep inelastic scattering (DIS).

Both inclusive ($\gamma^*p\!\to\!X$) and diffractive ($\gamma^*p\!\to\!Xp$)
DIS are processes in which a photon (of virtuality $Q^2$) is used as the hard probe,
and at small values of $x\!\simeq\!Q^2/W^2,$ parton saturation becomes relevant.
The dipole picture naturally describes inclusive and diffractive events within the same theoretical framework. It expresses the scattering of the virtual photon through its fluctuation into a color singlet $q\bar q$ pair (or dipole) of a transverse size $r\!\sim\!1/Q$. The dipole is then what probes the target proton, seen as a Color Glass Condensate (CGC): a dense system of gluons that interact coherently. Therefore, despite its perturbative size, the dipole cross-section is comparable to that of a pion. 

The same dipole scattering amplitude $\langle T(r)\rangle_x$ enters in the formulation of the inclusive and diffractive cross-sections:
\vspace{-0.5cm}\begin{eqnarray}
&&\scriptstyle\hspace{0.25cm}r<1/Q\hspace{1cm}1/Q<r<1/Q_s\hspace{1cm}r>1/Q_s\nonumber\\\frac{Q^2}{Q_s^2}\frac{d\sigma_{tot}}{d^2b}=&\displaystyle
4\pi \frac{Q^2}{Q_s^2}\int_0^\infty rdr\ \phi(r,Q^2)\langle T(r)\rangle_x
\simeq&\hspace{0.5cm}1\hspace{0.7cm}+\hspace{0.5cm}\ln\left(\frac{Q^2}{Q_s^2}\right)
\hspace{0.5cm}+\hspace{0.5cm}1\nonumber\\
\frac{Q^2}{Q_s^2}\frac{d\sigma_{diff}}{d^2b}=&\displaystyle
2\pi \frac{Q^2}{Q_s^2}\int_0^\infty rdr\ \phi(r,Q^2)\langle T(r)\rangle^2_x
\simeq&\hspace{0.4cm}\frac{Q_s^2}{Q^2}\hspace{0.5cm}+\hspace{1.1cm}1\hspace{1.1cm}+\hspace{0.5cm}1\nonumber
\end{eqnarray}
where $\phi(r,Q^2)$ is the well-known $\gamma^*\!\to\!q\bar q$ wavefunction. To obtain the right-hand sides, we have decomposed the dipole-size integration into three domains: $r\!<\!1/Q,$ $1/Q\!<\!r\!<\!1/Q_s,$ and $r\!>\!1/Q_s,$ and used the dipole amplitude $\langle T(r)\rangle_x$ discussed below. One sees that hard diffractive events
($Q^2\!\gg\!Q_s^2$) are much more sensitive to saturation than inclusive events, as the contribution of small dipole sizes is suppressed and the dominant size is 
$r\!\sim\!1/Q_s.$

\section{Hard diffraction off a Color Glass Condensate}

The Good-and-Walker picture of diffraction was originally meant to describe soft diffraction. They express an hadronic projectile 
$|P\rangle\!=\!\sum_n c_n|e_n\rangle$ in terms of hypothetic eigenstates of the 
interaction with the target $|e_n\rangle,$ that can only scatter elastically:
$\hat{S}|e_n\rangle\!=\!(1\!-\!T_n)|e_n\rangle.$ The total, elastic and 
diffractive cross-sections are then easily obtained:
\begin{equation}
\sigma_{tot}=2\sum_n c_n^2 T_n\hspace{0.5cm}
\sigma_{el}=\Big[\sum_n c_n^2 T_n\Big]^2\hspace{0.5cm}
\sigma_{diff}=\sum_n c_n^2 T_n^2\ .\label{gaw}
\end{equation}
 
It turns out that in the high energy limit, there exists a basis of eigenstates 
of the large$-N_c$ QCD $S-$matrix: sets of quark-antiquark color dipoles 
$|e_n\rangle\!=\!|d(r_1),\dots,d(r_n)\rangle$ characterized by their transverse 
sizes $r_i.$ In the context of deep inelastic scattering (DIS), we also know the 
coefficients $c_n$ to express the virtual photon in the dipole basis. For 
instance, the equivalent of $c_1^2$ for the one-dipole state is the photon wavefunction $\phi(r,Q^2).$ 

\begin{wrapfigure}{r}{0.5\columnwidth}
\centerline{\includegraphics[width=0.45\columnwidth]{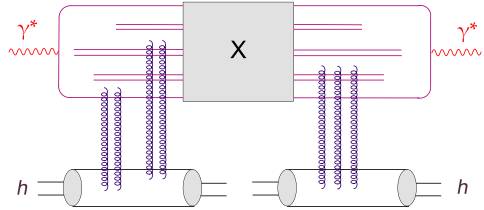}}
\caption{Representation of the factorization formula (\ref{diff}) for the 
diffractive cross-section in DIS. The virtual photon is decomposed into dipoles
which interact elastically with the target hadron. The rapidity gap is $Y_g$ and 
the final state $X$ is made of particles produced over a rapidity interval 
$Y-Y_g.$}\label{DDIS}
\end{wrapfigure}

This realization of the Good-and-Walker picture allows to write down exact 
(within the high-energy and large$-N_c$ limits) factorization 
formulae for the total and diffractive cross-sections in DIS. They are expressed in terms of elastic scattering amplitudes of dipoles off the CGC
$\left\langle T_n(\{r_i\})\right\rangle_Y,$ where $Y\!=\!\ln(1/x)$ is the total rapidity. The average $\langle\ .\ \rangle_Y$ is an average over the CGC wavefunction that gives the energy dependence to the cross-sections.

Formulae are similar to (\ref{gaw}) with extra integrations over the dipoles 
transverse coordinates. For instance, denoting the minimal rapidity gap $Y_g,$ the diffractive cross-section reads~\cite{difscal}
\begin{equation}
\sigma_{diff}(Y,Y_g,Q^2)=\sum_n\int dr_1\cdots dr_n\ 
c_n^2(\{r_i\},Q^2,Y\!-\!Y_g)\  \left\langle T_n(\{r_i\})\right\rangle_{Y_g}^2\ .\label{diff}
\end{equation}
This factorization is represented in Fig.~\ref{DDIS}. Besides the $Q^2$ dependence, the probabilities to express the virtual photon in the dipole basis $c^2_n$ also depend on $Y\!-\!Y_g.$ Starting with the initial condition 
$c_n^2(\{r_i\},Q^2,0)\!=\!\delta_{1n}\phi(r,Q^2),$ the probabilities can be 
obtained from the high-energy QCD rapidity evolution. Finally, the scattering 
amplitude of the n-dipole state $T_n(\{r_i\})$ is given by
\begin{equation}
T_n(\{r_i\})=1-\prod_{i=1}^n(1-T(r_i))\nonumber
\end{equation}
where $T(r)\!\equiv\!T_1(r)$ is the scattering amplitude of the one-dipole state. The rapidity evolution of the correlators 
$\left\langle T(r_1)\dots T(r_n)\right\rangle_Y$ should obtained from the CGC 
non-linear equations; one can then compute the diffractive cross-section.

When taking $Y_g\!\to\!Y$ in formula \eqref{diff}, one recovers the formula used for our previous estimates, which corresponds to restricting the diffractive final state to a $q\bar q$ pair. In practice the description of HERA data also requires a 
$q\bar qg$ contribution.

\section{The CGC non-linear evolution equations}

Within the high-energy and large$-N_c$ limits, the scattering amplitudes off the CGC are obtained from the Pomeron-loop equation~\cite{ploop} derived in the 
leading logarithmic approximation in QCD. This is a Langevin equation which exhibits the stochastic nature~\cite{fluc} of high-energy scattering processes in QCD. Its solution $\bar{T}$ is an event-by-event dipole scattering amplitude function of 
$\rho\!=\!-\ln(r^2Q_0^2)$ and $Y$ ($Q_0$ is a scale provided by the initial condition). 

\begin{wrapfigure}{r}{0.5\columnwidth}
\centerline{\includegraphics[width=0.45\columnwidth]{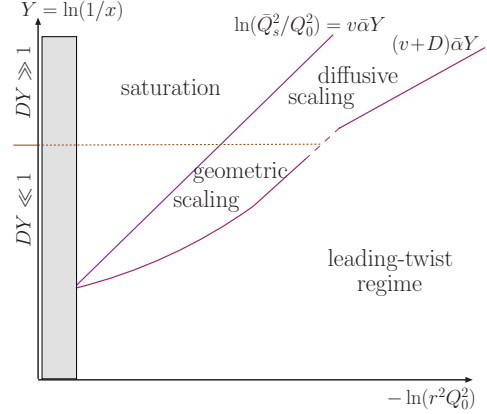}}
\caption{A diagram summarizing the high-energy QCD non-linear evolution. Shown are the average saturation line and the boundaries of the scaling regions at small values of $r.$ With increasing $Y,$ there is a gradual transition from geometric scaling at intermediate energies to diffusive scaling at very high energies.}
\label{ploop}
\end{wrapfigure}

The solution $\bar{T}(\rho,Y)$ is characterized by a saturation scale $Q_s$ which is a random variable whose logarithm is distributed according to a Gaussian probability law~\cite{prob}. The average value is 
$\ln(\bar{Q}_s^2/Q_0^2)\!=\!v\bar\alpha Y$ and the variance is 
$\sigma^2\!=\!D\bar\alpha Y.$ The saturation exponent $v$ determines the growth of 
$\bar{Q}_s$ with rapidity, and the dispersion coefficient 
$D$ defines two energy regimes: the geometric scaling regime ($DY\!\ll\!1$) and diffusive scaling regime ($DY\!\gg\!1$).

Evolving a given initial condition yields a stochastic ensemble of solutions
$\bar{T},$ from which one obtains the dipole correlators:
\begin{equation}
\left\langle T(r_1)\dots T(r_n)\right\rangle_Y\!=\!
\left\langle \bar{T}(\rho_1,Y)\dots \bar{T}(\rho_n,Y)\right\rangle
\nonumber\end{equation}
where in the right-hand side, $\langle\ .\ \rangle$ is an average over the realizations of $\bar{T}.$ Indeed, both quantities $\left\langle T\dots T\right\rangle_Y$ and $\left\langle \bar{T}\dots \bar{T}\right\rangle$ obey the same hierarchy of equations. One obtains the following results for the dipole scattering amplitudes~\cite{errorf}:
\begin{eqnarray}
\left\langle T(r_1)\dots T(r_n)\right\rangle_Y&\stackrel{Y\ll1/D}{=}&
\left\langle T(r_1)\right\rangle_Y\dots\left\langle T(r_n)\right\rangle_Y\ ,
\nonumber\\
\left\langle T(r_1)\dots T(r_n)\right\rangle_Y&\stackrel{Y\gg1/D}{=}&
\left\langle T(r_<)\right\rangle_Y,\hspace{0.3cm}r_<=\min(r_1,\dots,r_n)
\nonumber\ .
\end{eqnarray}
All the scattering amplitudes are expressed in terms of $\langle T(r)\rangle_Y,$ 
the amplitude for a single dipole which features the following scaling 
behaviors:
\begin{eqnarray}
\left\langle 
T(r)\right\rangle_Y\stackrel{Y\ll1/D}{\equiv}T_{gs}(r,Y)&=&\displaystyle
T\left(r^2\bar{Q}_s^2(Y)\right)\ ,\label{gs}\\
\left\langle 
T(r)\right\rangle_Y\stackrel{Y\gg1/D}{\equiv}T_{ds}(r,Y)&=&\displaystyle
T\left(\frac{\ln(r^2\bar{Q}_s^2(Y))}{\sqrt{DY}}\right)\ .\label{ds}
\end{eqnarray}

In the saturation region $r\bar{Q}_s\!>\!1,$ $\left\langle 
T(r)\right\rangle_Y\!=\!1.$ As the dipole size $r$ decreases, $\left\langle 
T(r)\right\rangle_Y$ decreases towards the weak-scattering regime following the 
scaling laws (\ref{gs}) or (\ref{ds}), depending on the value of $DY$ as shown in Fig.~\ref{ploop}. In the geometric scaling regime ($DY\!\ll\!1$), the dispersion of the events is negligible and the averaged amplitude obeys (\ref{gs}). In the diffusive scaling regime ($DY\!\gg\!1$), the dispersion of the events is important, resulting in the behavior (\ref{ds}). When Pomeron loops are not included in the evolution, only the geometric scaling regime appears.

\section{Phenomenology}

In the geometric scaling regime, instead of being a function of the two variables 
$r$ and $x,$ $T_{gs}(r,Y)$ is a function of the single variable $r\bar{Q}_s(x)$ up to inverse dipole sizes significantly larger than the saturation scale 
$\bar{Q}_s(x).$ This means that in the geometric scaling window in 
Fig.~\ref{ploop}, $T_{gs}(r,Y)$ is constant along lines parallel to the saturation line. Physically, they are lines along which the dipole sees a constant partonic density inside the proton. 

In DIS, this feature manifests itself via the so-called geometric scaling property. Instead of being a function of $Q^2$ and $x$ separately, the total cross-section is only a function of $\tau\!=\!Q^2/\bar{Q}_s^2(x),$ up to large values of $\tau;$ similarly, the diffractive cross-section is only a function of 
$\tau_d\!=\!Q^2/\bar{Q}_s^2(x_{\mathbb P}),$ and $\beta:$ 
\begin{eqnarray}
\sigma^{\gamma^*p\rightarrow X}_{tot}(x,Q^2)&=&
\sigma^{\gamma^*p\rightarrow X}_{tot}(\tau\!=\!Q^2/\bar{Q}_s^2(x))\ ,
\nonumber\\
\sigma^{\gamma^*p\rightarrow Xp}_{diff}(\beta,x_{\mathbb P},Q^2)&=&
\sigma^{\gamma^*p\rightarrow Xp}_{diff}
(\beta,\tau_d\!=\!Q^2/\bar{Q}_s^2(x_{\mathbb P}))\nonumber\ .
\end{eqnarray}
Experimental measurements are compatible with those predictions~\cite{gs}, with the parameters $\lambda\!\simeq\!0.25$ and $x_0\!\simeq\!10^{-4}$ for the average saturation scale $\bar{Q}_s(x)=(x/x_0)^{-\lambda/2}\ \mbox{GeV}.$ This determines the saturation exponent $v\!=\!\lambda/\bar\alpha.$ 
HERA probes the geometric scaling regime and one could expect so of future measurements at an electron-ion collider.

The estimates of Section I (where one should now replace $Q_s$ by $\bar{Q}_s$) are obtained in the geometric scaling regime: the total cross-section is dominated by semi-hard sizes ($1/Q\!<\!r\!<\!1/\bar{Q}_s$) while the diffractive cross-section is dominated by dipole sizes of the order of the hardest infrared cutoff in the 
problem: $1/\bar{Q}_s.$ In the diffusive scaling regime, up to values of $Q$ much bigger than the average saturation scale $\bar{Q}_s,$ things change drastically:
both inclusive and diffractive scattering are dominated by small dipole sizes, of order $1/Q,$ yet saturation plays a crucial role. Cross-sections are dominated by rare events in which the photon hits a black spot that he sees at saturation at the scale $Q^2.$ In average the scattering is weak ($T_{ds}(r,Y)\!\ll\!1$), but saturation is the relevant physics. 

Our poor knowledge of the coefficient $D$ prevents quantitative analysis, still the diffusive scaling regime has striking signatures. For instance the inclusive and diffractive cross-sections do not feature any Pomeron-like (power-law type) increase with the energy. It is likely out of the reach of HERA, and future studies in the context of $p\!-\!p$ collisions at the LHC are certainly of interest.


\begin{footnotesize}



%

\end{footnotesize}


\end{document}